\documentclass{osa-article}

\journal{osajournal}

\articletype{Research Article}

\usepackage{lineno}

\begin{document}

\title{Computational 3D topographic microscopy from terabytes of data per sample}

\author{Kevin C. Zhou,\authormark{1,2,4,}* Mark Harfouche,\authormark{2} Maxwell Zheng,\authormark{2} Joakim Jönsson\authormark{1}, Kyung Chul Lee\authormark{1,3}, Ron Appel\authormark{2}, Paul Reamey,\authormark{2} Thomas Doman,\authormark{2} Veton Saliu,\authormark{2} Gregor Horstmeyer,\authormark{2} and Roarke Horstmeyer\authormark{1,2,}*}

\address{\authormark{1}Department of Biomedical Engineering, Duke University, Durham, NC 27708, USA\\
\authormark{2}Ramona Optics Inc., 1000 W Main St., Durham, NC 27701, USA\\
\authormark{3}School of Electrical and Electronic Engineering, Yonsei University, Seoul, 03722, South Korea\\
\authormark{4}Current affiliation: Department of Electrical Engineering and Computer Sciences, University of California, Berkeley, CA, USA\\
\authormark{*}\email{kevinczhou@berkeley.edu} \authormark{*}\email{roarke.w.horstmeyer@duke.edu}
}




\begin{abstract*}
We present a large-scale computational 3D topographic microscope that enables 6-gigapixel profilometric 3D imaging at micron-scale resolution across $>$110 cm\textsuperscript{2} areas over multi-millimeter axial ranges. Our computational microscope, termed STARCAM (Scanning Topographic All-in-focus Reconstruction with a Computational Array Microscope), features a parallelized, 54-camera architecture with 3-axis translation to capture, for each sample of interest, a multi-dimensional, 2.1-terabyte (TB) dataset, consisting of a total of 224,640 9.4-megapixel images. We developed a self-supervised neural network-based algorithm for 3D reconstruction and stitching that jointly estimates an all-in-focus photometric composite and 3D height map across the entire field of view, using multi-view stereo information and image sharpness as a focal metric. The memory-efficient, compressed differentiable representation offered by the neural network effectively enables joint participation of the entire multi-TB dataset during the reconstruction process. To demonstrate the broad utility of our new computational microscope, we applied STARCAM to a variety of decimeter-scale objects, with applications ranging from cultural heritage to industrial inspection.
\end{abstract*}

\section{Introduction}
All optical imaging systems operate within a trade-off space, in which resolution, field-of-view (FOV), and imaging speed must all be carefully selected for a given application of interest. For example, due to both practical and physics-based constraints, widely used commercial and high-end objective lenses can only resolve a limited number of points within their FOV \cite{park2021review}. This problem tends to get worse at higher spatial resolutions \cite{zheng2014fourier}, as higher-order aberrations become practically more difficult and expensive to correct over wide FOVs. Furthermore, as lateral resolution increases, an imaging system's depth of field (DOF) becomes quadratically narrower, diminishing the axial FOV and therefore the number of axially resolvable points for 3D imaging applications. As a result, many imaging techniques designed to capture 3D surface profiles, such as photogrammetry \cite{luhmann2010close}, active stereo \cite{scharstein2003high}, structured light imaging \cite{yen2006full,geng2011structured}, and line structured light imaging \cite{xu2020line}, have largely been applied in lower-resolution, low-magnification applications and offer macroscopic FOVs (centimeter and decimeter-scale). There are very few imaging methods that can acquire high, microscopic resolution 3D surface measurements across large areas, which is the central goal of this work.

The ability to jointly measure the 3D properties of large, macroscopic surfaces at high resolution can benefit a variety of applications. Such an instrument could, for example, fully digitize macroscopic 3D objects at microscopic resolution, which would prove valuable within cultural heritage, in particular for the inspection and digitization of artwork \cite{spagnolo2000three,pieraccini20013d}. Such a method would also find value within the industrial inspection of electronics components\cite{traxler2021experimental}, including wafer defect detection \cite{wu2014wafer}, printed circuit boards (PCBs) \cite{acciani2006application,yen2006full,hong2009sensor}, and chip-scale packages (CSPs) \cite{xue2009warpage,li2014telecentric}. While there are relevant approaches that push higher-resolution, micron-scale profilometry, such as microscopic structured light imaging \cite{hu2020microscopic} and interferometric techniques \cite{chen20103,czajkowski2010optical} generally operate over smaller FOVs (millimeter-scale). These existing methods therefore require the slow process of scanning and tiling, one snapshot at a time, to achieve high resolution across sufficiently large FOVs, including axial scanning to overcome the impact of a limited DOF at high resolution. There is thus a need for a high-resolution topographic imaging system that can acquire data at reasonable speeds across large, macroscopic objects, along with new computational strategies that can scalably handle the associated orders-of-magnitude-increased dataset sizes, which could open up a wide range of exciting applications surrounding the topics outlined above.

\begin{figure}
    \centering
    \centerline{\includegraphics[width=1.2\textwidth]{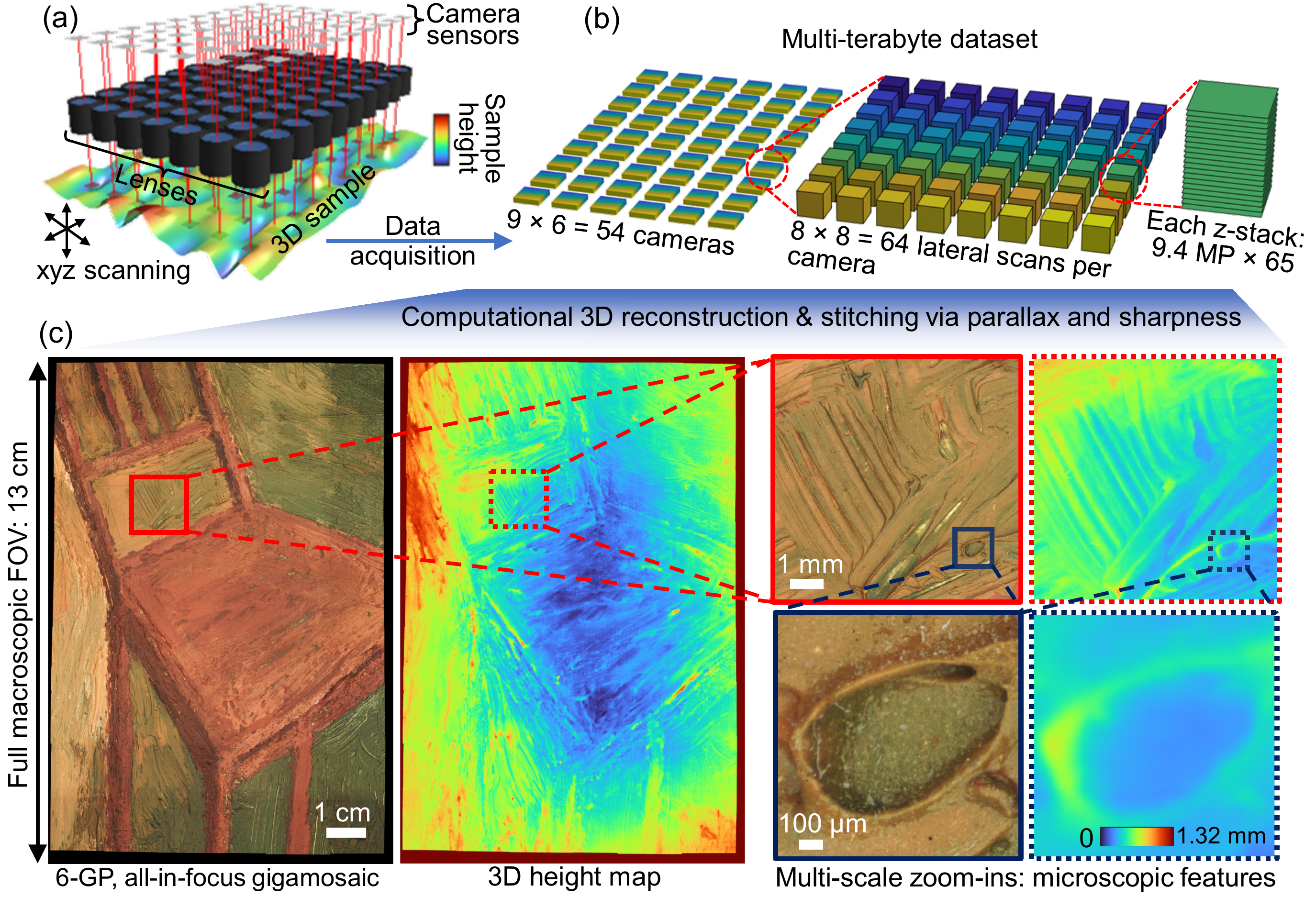}}
    \caption{Overview of STARCAM. (a,b) Data acquisition involves scanning a 3D object in three dimensions and synchronously capturing $\sim$9.4-MP, high-resolution images across an array of 9$\times$6 cameras. Lateral scanning (8$\times$8) is necessary to fill in the gaps between sensors, due to the high magnification, as well as provide stereo information for 3D estimation. z-stacking (65 steps) enables 3D topographic estimation of thicker samples using sharpness measures. (c) Our computational 3D reconstruction algorithm generates a 6-GP, all-in-focus gigamosaic along with a coregistered 3D height map.}
    \label{fig:teaser}
\end{figure}

To this end, to overcome the current 3D measurement throughput limitations of existing 3D profilometric techniques, we present a parallelized computational 3D topographic microscope that can perform 3D surface profilometric imaging at micron-scale resolution over a 13 cm $\times$ 9 cm$>$110 cm\textsuperscript{2} lateral FOV and multi-millimeter axial ranges (Fig. \ref{fig:teaser}). Our method, termed STARCAM (\underline{S}canning \underline{T}opographic \underline{A}ll-in-focus \underline{R}econstruction with a \underline{C}omputational \underline{A}rray \underline{M}icroscope), uses a multi-camera array microscope (MCAM \cite{thomson2022gigapixel,harfouche2023imaging,zhou2023parallelized,yang2022multi}) and 3-axis sample scanning to capture a 2.1-terabyte (TB) dataset for each sample of interest. Our MCAM contains a 9$\times$6 array of cameras, in principle allowing for 54$\times$ increased throughputs beyond conventional microscopes. Notably, the range of sample scanning is limited to the inter-camera spacing rather than the total extended FOV, and the total scan time is independent of the number of cameras. We then process the multidimensional, multi-TB dataset through a large-scale, self-supervised, neural-network-based reconstruction and stitching algorithm that estimates an all-in-focus (AiF) 6-gigapixel (GP) photometric composite along with a coregistered 3D height map, using the stereo cues from the overlapped lateral scanning and sharpness cues from the axial scanning. The neural network acts as a compressed differentiable representation of the reconstruction that enables a memory-efficient computational reconstruction process, while still effectively allowing joint participation of the entire multi-TB dataset by loading and preprocessing random z-stack patches from storage to computer memory on the fly. 

We applied STARCAM to a wide variety of decimeter-scale 3D objects, including an oil painting, PCB, and multiple CPU pin grid arrays (PGAs) and ball grid arrays (BGAs) in parallel. Our multi-GP 3D topographic microscopy technology paves the way to a solution to the high-throughput imaging demand of future industrial inspection applications that require in-line monitoring of complex parts at multiple stages of fabrication.

\section{Multi-terabyte data acquisition for high-SBP 3D topographic microscopy}

\subsection{Multi-camera array microscope (MCAM) hardware design}
The MCAM's highly parallelized design consists of an array of 9$\times$6=54  micro-camera units, spaced by 1.35 cm in both lateral dimensions, each with a 13-MP Bayered CMOS sensor (3120$\times$4208, pixel size=1.1\textmu m) \cite{harfouche2023imaging}. For all experiments in this study, we used $3072\times3072$ square crops.
Each micro-camera is equipped with a 25.05-mm effective focal length (EFL) lens (Edmund Optics), axially positioned to form a finite-conjugate, non-telecentric imaging configuration with a magnification of $\sim$0.8$\times$ (corresponding an object-side digital resolution of 1.378\textmu m) and an object-side numerical aperture (NA) of $\sim$0.088. The sample is illuminated using white LEDs surrounding the micro-camera apertures.

\subsection{Multi-terabyte data acquisition}
The inter-micro-camera spacing is larger than the per-micro-camera FOV (4.1 mm), leading to gaps in the extended FOV for a single synchronized snapshot. Furthermore, the sample height variation typically extends beyond the system depth of field (DOF). To fill in these lateral gaps and to cover an extended axial range, we placed the samples on a 3-axis translation stage (Zaber X-LSM) and translated them laterally in an 8 by 8 $\mathit{xy}$ grid, and axially across 65 $z$ planes, spanning up to 4.8 mm for the samples analyzed in this paper. Note that the stage only needs to travel laterally less than the inter-camera spacing (1.35 cm) and not the entire FOV, resulting in faster acquisition times compared to single-aperture systems covering the same area. In particular, we fixed $\mathit{xy}$ translation step size to 13.5/8 mm = 1.6875 mm, resulting in $>$50\% overlap redundancy between adjacent scans to provide stereo information to facilitate 3D estimation. 

In sum, for each sample, we capture synchronized snapshots across up to 54 cameras across $8\times8\times65$ scan positions, resulting in a 7D data hypervolume, with dimensions of $9\times6\times8\times8\times65\times3072\times3072$, corresponding to 224,640 9.4-MP images, or 2.1 TB of raw data per sample (saved as unsigned 8-bit integers, with a single bayered channel). The MCAM is able to stream data at $>$5 GB/sec, theoretically enabling capture of the entire 7D hypervolume in only 7 minutes. In practice, our data acquisition speeds are limited by the stage translation speed and settling time. The acquisition speed could also be further increased using sample-adaptive strategies, such as with tunable-focus lenses.

\section{Mechanisms for 3D estimation: multi-view stereo and height from sharpest focus}\label{3d_mechanism}

\begin{figure}[h]
    \centering
    \centerline{\includegraphics[width=1.3\textwidth]{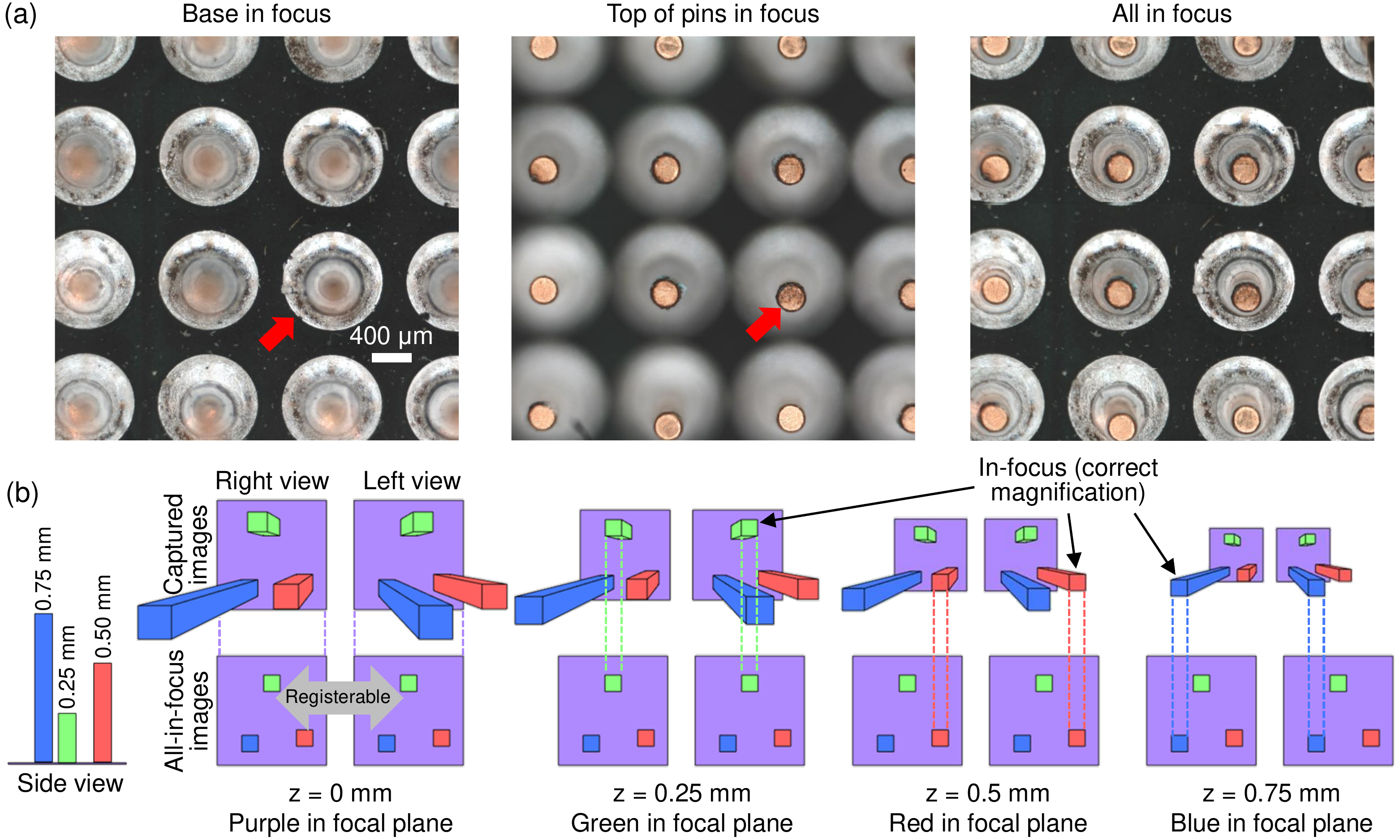}}
    \caption{The two mechanisms for 3D estimation. (a) Image sharpness can be used to estimate height. The left image shows the base of a pin grid array (PGA) in focus, while the middle image shows the tips of the pins of the PGA in focus. The right image shows the all-in-focus reconstruction. (b) Height can also be estimated by eliminating parallax. Only when a feature is at the nominal focal plane of the imaging system does it have the correct magnification and therefore in the correct location (no parallax). When the feature is out of focus, it moves away from or towards the optical axis (parallax). When all features are in focus, the images from different cameras are directly registerable using homographies. }
    \label{fig:3d_mechanism}
\end{figure}

Since we translated the sample axially and ensured $>$50\% lateral overlap across the entire FOV, we were able to take advantage of two different methods for 3D estimation simultaneously, 3D from stereo vision \cite{furukawa2015multi} and sharpest focus \cite{krotkov1988focusing}, even though the experimental requirements of these two methods generally conflict. In particular, while stereo techniques require long DOFs, which determine the axial range of 3D estimation, sharpness-based approaches work best with narrow DOFs, which dictate the axial resolution of 3D estimation. Furthermore, while stereo techniques typically use a pinhole camera model, which implicitly assumes non-telecentric optics, sharpness-based approaches work best used with telecentric optics (or make a telecentric approximation over a restricted FOV). Despite these seemingly disparate requirements, we show that our method is able to take advantage of both physical mechanisms synergistically for robust 3D estimation at a few-micron lateral resolution over $>$10 cm FOVs.

\subsection{Height from sharpest focus}\label{height_from_sharpest_focus}
Estimating depth from z-stacks using a sharpness metric applied slice-wise is perhaps most intuitive and straightforward with telecentric optics, which guarantee a constant, depth-independent magnification. This property ensures that every feature within the lateral FOV stays in the same place during z-stack acquisition, thereby enabling direct sharpness comparisons. Thus, the argmax operation (i.e., the position of the largest value) across the $z$ dimension applied pixel-wise would yield a good estimate of the sample height map. However, telecentric optics are disadvantageous in that they restrict the lateral FOV (i.e., to less than the diameter of the lens).

Perhaps counterintuitively, even for z-stacks captured with non-telecentric imaging configurations (as in our MCAM design), the same argmax operation would still in principle yield an accurate height estimate. That is, even though the z-stack exhibits depth-dependent magnification, the individual object features will be the sharpest when they intersect with the focal plane and thus will be governed by a common magnification (Fig. \ref{fig:3d_mechanism}b). The difference from the telecentric case is that in the non-telecentric case, except at the very center of the FOV, the argmax operation must be able to identify the sharpest object features among blurred versions of other object features from different lateral positions. Thus, the sharpness metric should ideally be robust to changes in object appearance (see the weighted sharpness loss in Sec. \ref{self_supervised}). Incidentally, correcting for depth-dependent magnification changes to enable more direct sharpness comparisons would essentially be tantamount to estimating the height map. We thus additionally use stereo cues to improve 3D estimation.

\subsection{Height from stereo}

As mentioned earlier, the data acquisition procedure ensured at least 50\% overlap in both lateral dimensions. Thus, any point apart from those on the outer edges are viewed from at least four different perspectives. This multi-view information provides stereo parallax cues for 3D estimation \cite{furukawa2015multi,zhou2021mesoscopic,zhou2023parallelized}. Note that in combination with the z-stacking, we overcome the implicit resolution limits of stereo techniques, imposed by their requirement for long DOFs. The way stereo information is incorporated is by enforcing consistency in the height values predicted based on the sharpness cues in overlapping regions. In particular, since each z-stack results in radial expansion about its respective center due to depth-dependent magnification, in overlapped regions of neighboring views the expansions occur in opposite directions. This is known as parallax. While our previous plane-plus-parallax implementations use orthorectification \cite{zhou2021mesoscopic,zhou2023parallelized} (i.e., undo these radial shifts due to depth-dependent magnification) by deforming a single image, in our case we retrieve the correct pixel from the corresponding depth slice of the z-stack.

\section{3D topographic reconstruction and stitching algorithm}
The STARCAM computational 3D topographic reconstruction algorithm extends our previous self-supervised 3D-RAPID algorithm \cite{zhou2023parallelized} to much larger spatial scales, ranging from 40$\times$ to 650$\times$ larger SBPs per frame and 3-4 orders of magnitude more raw data per frame. Like 3D-RAPID, our algorithm jointly reconstructs across the entire FOV, requiring the simultaneous participation of z-stacks across all micro-cameras and all scan positions. This strategy promotes a globally-consistent reconstruction that would not be possible with a sequential algorithm. As such, we had to develop new data handling strategies, especially considering that the 8-bit integers need to be converted to 32-bit floats for computation, which would cause the 2.1-TB dataset size to balloon up to 8.4 TB and therefore easily exceed the RAM capacity of even the highest-end of current consumer computers. Furthermore, it is non-ideal to make copies of the dataset refactored for batched training, due to the increased data storage requirements.

\subsection{Joint camera and sample scan calibration: 6D pose, distortion, intensity variation, and focal plane shift}\label{camera_calibration}

To maximize registration, stitching, and 3D estimation accuracy, we pre-calibrated the microscope by capturing a 7D hypervolume of a flat patterned target. Note that a z-stack is still necessary for a flat target because the focal planes of the 54 micro-cameras in practice may not be coincident. For each z-stack, we pick the sharpest slice (using a Laplacian-based metric) and jointly registered the $9\times6\times8\times8=3456$ images by simultaneously optimizing the 6D poses (3D position + 3D orientation) for all 3456 images, along with a quadratic radial distortion parameter, assumed to be the same for all images, and inter-image and intra-image intensity variation (e.g., due to uneven illumination). The pixel-intensity-based joint registration algorithm extends our previously reported algorithm \cite{zhou2021mesoscopic} by incorporating new strategies to support reconstructions with much higher SBPs, as required by the large datasets acquired for this study. 

Before we discuss the new extensions, we briefly review the base algorithm \cite{zhou2021mesoscopic}, which takes in a collection of images to be registered according to an image deformation model using gradient descent. For the case of camera pose and distortion calibration, the deformation model is the 6D pose modeling homographic distortions, along with distortion parameters. All the images are dewarped according to an initial guess of the deformation parameters and projected onto a blank canvas to form an estimate of the registered composite. When a pixel in the composite is visited multiple times (i.e., a collision), the values are averaged. To quantify how accurate the deformation parameters are, the values are reprojected from the composite back to the image space and compared to the original images via mean square error (MSE). This MSE is minimized with respect to the deformation parameters via gradient descent, reaching a minimum when all the pixel collisions are consistent. Note that the reconstruction is reset at every gradient descent iteration.

This algorithm has worked well for jointly calibrating 10s of multi-MP cameras that form composites of up to a few 100s of MP. To extend this algorithm to jointly register 3456 images to form a 6-GP composite, we introduced a multi-scale strategy along with pixel batching. In particular, first we optimize the deformation model parameters for the data captured by a single micro-camera (i.e., 8$\times$8 images of size 3072$\times$3072), using the previous method. An example registration is visualized in Supplementary Fig. \ref{fig:calibration_overlap}a. The optimized parameters for the one micro-camera are then used to initialize the parameters for all micro-cameras, with a rough initial guess for the inter-camera spacing. From there, all 3456 camera images are coarsely registered with 53$\times$ linear downsampling, updating only the lateral camera positions during gradient descent. Next, the downsampling is decreased to 26$\times$ and the full 6D poses for all 3456 images are simultaneously updated. 

Finally, the downsampling is decreased to 4$\times$, which makes the multi-view image dataset too big for the previous approach. To overcome this challenge, instead of having all pixels from all images contributing to gradient descent, we select a random batch of pixels across all images at each gradient descent iterations. Here, the reconstruction is not reset after every iteration, but rather is updated as a moving average across sequential batches. This moving average image registration approach is similar to our previous algorithms \cite{zhou2021mesoscopic, zhou2022computational}, except the batches are random pixels rather than random multi-view images or image columns. In this final step, all calibration parameters are optimized. An example registration of all 3456 images is visualized in Supplementary Fig. \ref{fig:calibration_overlap}b.

\subsection{Joint reconstruction of 3D topography across the entire FOV}
\begin{figure}
    \centering
    \centerline{\includegraphics[width=1.4\textwidth]{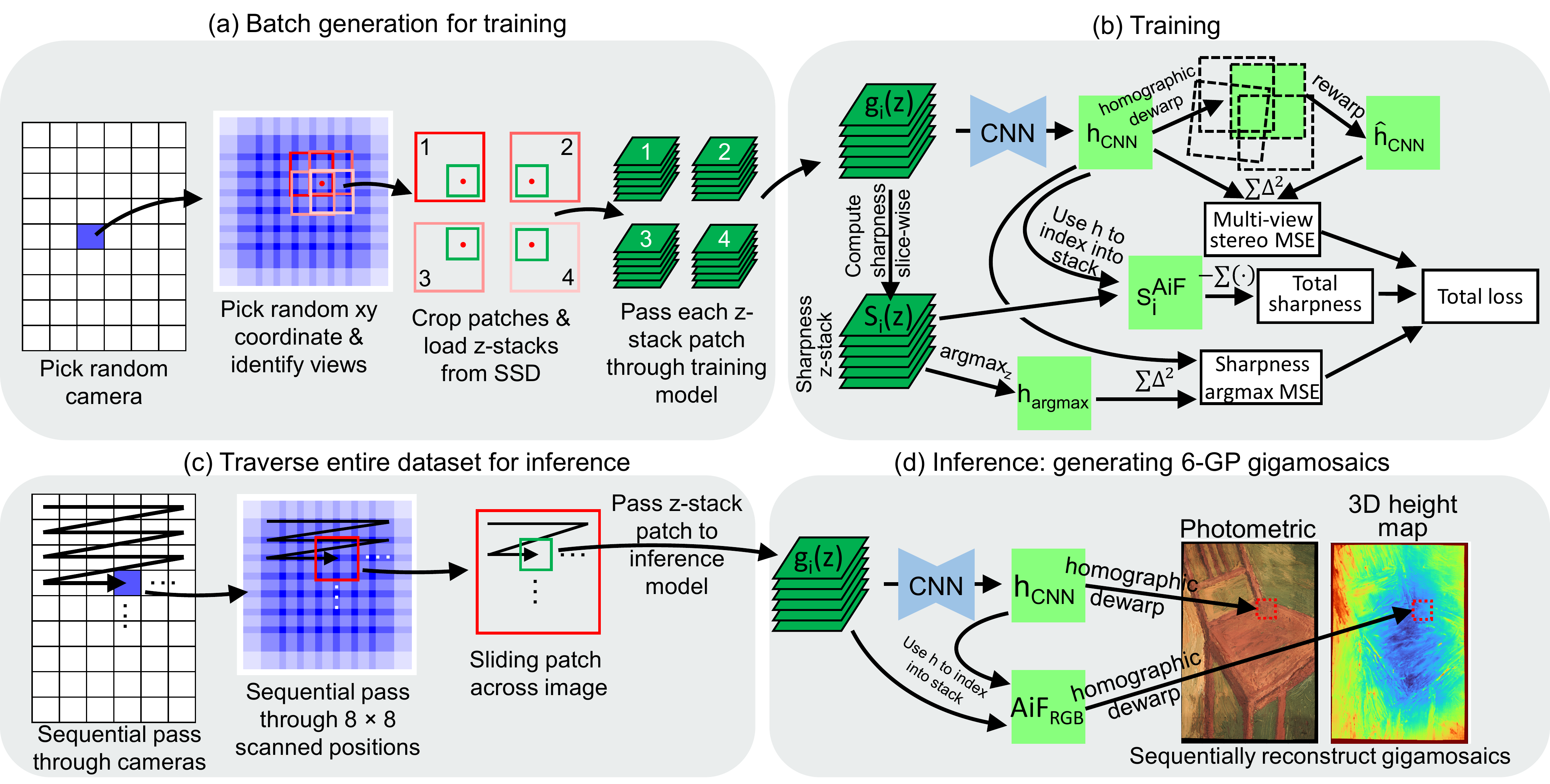}}
    \caption{Computational 3D topographic and all-in-focus (AiF) self-supervised reconstruction algorithm, using stereo and focus cues for supervision. (a) Procedure for generating a single batch element, consisting of a collection of 2-9 z-stack patches (the case of 4 illustrated here). (b) The batches of z-stack patches are passed through the computational model to generate a scalar loss to be minimized by gradient descent. (c)-(d) CNN inference for generating the 6-GP photometric and 3D height map gigamosaics.}
    \label{fig:algorithm}
\end{figure}

Once the cameras are calibrated, the goal is to reconstruct an all-in-focus (AiF) photometric (RGB) gigamosaic along with a coregistered 3D height map. To do this in a compact and memory-efficient manner, we optimize a CNN in an end-to-end fashion to map from z-stacks to the 3D height map via self-supervised learning. Our training procedure does not require any data beyond the 7D hypervolume for the sample of interest, with the supervision coming from the stereo and height-from-sharpest-focus cues (Sec. \ref{3d_mechanism}). In other words, the 3D reconstruction procedure itself is the training procedure of the CNN, which acts as a compact representation of the object's photometric and height properties. This reparameterization leads to a memory-efficient reconstruction algorithm because the decompressed 6-GP photometric and 3D height map gigamosaics never materialize until after training, during CNN inference. Furthermore, the CNN confers regularizing effects due to their inductive biases \cite{ulyanov2018deep} and compressive representation (i.e., the CNN contains fewer parameters than the number of pixels in the gigamosaics). The CNN architecture is identical to the one used in 3D-RAPID \cite{zhou2023parallelized}, except that the color channel of the CNN input is replaced with the z-stack dimension (and only the debayered green channel is used).

\subsubsection{Self-supervised training}\label{self_supervised}
The self-supervised, patch-based CNN training algorithm of STARCAM is summarized in Fig. \ref{fig:algorithm}a, b. At each gradient descent iteration, random spatial patches of shape 576$\times$576 are hierarchically sampled by first picking a random camera (out of 54), followed by picking a random $xy$ coordinate from the FOV spanning the 8$\times$8 scan of that camera. All the images that capture that point, according to the camera calibration (Supplementary Fig. \ref{fig:calibration_overlap}) are identified, which can range from 2 to 9 images. Within each of these images, we identify the 576$\times$576 patch centered around that selected point. If the point is too close to the edge of the image, then we shift the patch until its within the image. Further, the pixel coordinates are rounded to the nearest even number to ensure correct alignment of the bayer pattern. Finally, we load from storage (NVMe SSD: Sabrent 4TB Rocket 4 Plus or Kioxia 7.68TB CD6-R) the 2-9 cropped z-stacks, corresponding to these patches, each with shape 576$\times$576$\times$65 -- this is the first time any sample data has been loaded. We then debayer each image and arbitrarily select only the green channel. Let the $i^\mathit{th}$ z-stack patch be denoted as $g_i(z)$. This procedure constitutes the generation of a single batch element. In our experiments, we use a batch size of 2. Currently, our optimization is bottlenecked by batch generation, taking around 2.5$\times$ longer to generate a batch on the CPU than for the GPU to consume it in a gradient update step. NVMe drives with faster read speeds will help bridge this gap. 

The batches of collections of z-stack patches ($\{g_i(z)\}_i$) is generated on the CPU, after which they are transferred to the GPU for the gradient descent step. Each z-stack patch is passed through the CNN, which predicts the height map, $h_\mathit{CNN}$ for that patch (Fig. \ref{fig:algorithm}b). We use three different loss functions to evaluate the fidelity of $h_\mathit{CNN}$ that encapsulate the two physical mechanisms for 3D estimation described in Sec. \ref{3d_mechanism}:
\begin{enumerate}
    \item \textbf{Stereo loss.} The height maps, $h_\mathit{CNN}$, corresponding to patches from different views are dewarped according to the precalibrated camera parameters (the homographic dewarp step shown in Fig. \ref{fig:algorithm}b), and then superimposed and averaged in overlapped regions. The patches are then reprojected back to camera-centric coordinates to form $\hat{h}_\mathit{CNN}$ for comparison with $h_\mathit{CNN}$ via MSE. Thus, minimizing this loss promotes accurate registration of the $h_\mathit{CNN}$ patches, enforcing stereo consistency between neighboring camera views. 
    \item \textbf{Weighted sharpness.} Each pixel of $h_\mathit{CNN}$ can be converted into a depth index to retrieve values from the original z-stack, $g_i(z)$ (indeed, this is how the AiF image is generated). However, instead of indexing into the original z-stack, we first compute a sharpness z-stack. To this end, we first divide each image in $g_i(z)$ by a Gaussian-blurred version ($\sigma$ = 8 pixels): 
    \begin{equation}
        g_i^\mathit{hpf}(z) = \frac{g_i(z)}{g_i(z)\circledast\mathit{Gauss}_\mathit{2D}}.
    \end{equation}
    This operation can be thought of as a normalized high-pass filter, to facilitate sharpness comparisons across different spatial regions due to depth-dependent magnification changes (Sec. \ref{height_from_sharpest_focus}). We then compute the magnitude of discrete spatial gradients, which we blur with the same Gaussian kernel,
    \begin{equation}\label{dxy_blur}
        S_i(z) = \left|\nabla_\mathit{x,y} g_i^\mathit{hpf}(z)\right|\circledast\mathit{Gauss}_\mathit{2D},
    \end{equation}
    which is the final sharpness metric. Ignoring padding issues, $S_i(z)$ and $g_i(z)$ have the same shape (i.e., 576$\times$576$\times$65). We then use $h_\mathit{CNN}$ to index into $S_i(z)$, generating $S_i^\mathit{AiF}(z)$. Note that to preserve differentiability, the indexing process using the height map interpolates between the two closest depth slices. Finally, we sum a weighted version of $S_i^\mathit{AiF}$ across lateral space to generate the final weighted sharpness loss, 
    \begin{equation}\label{weighted}
        \mathit{loss}_\mathit{sharpness} = \sum_{x,y} S_i^\mathit{AiF} \odot \mathit{max}\left( S_i^\mathit{AiF} - \delta, 0\right),
    \end{equation}
    where $\odot$ denotes element-wise multiplication, $max(\cdot,\cdot)$ is an element-wise maximum operation (equivalent to numpy's \texttt{maximum} function), and $\delta$ is a scalar constant hyperparameter. This loss gives higher weight to spatial regions that have high sharpness, and excludes low-sharpness regions from contributing to the loss (via $\delta$), which we empirically found to avoid artifacts in low-contrast samples (Supplementary Sec. \ref{confidence_maps}). We can also generate a confidence map based on the max sharpness across the axial dimension (Supplementary Sec. \ref{confidence_maps}).
    \item \textbf{Argmax.} We also apply the argmax operation to $S_i(z)$ (Eq. \ref{dxy_blur}) across $z$ to generate a height map estimate, $h_\mathit{argmax}$, which we compare with $h_\mathit{CNN}$ via MSE, weighted by the same factor as in Eq. \ref{weighted}. While $h_\mathit{argmax}$ sometimes produces artifacts, it provides long-distance gradients (i.e., when $h_\mathit{CNN}-h_\mathit{argmax}$ is large, the argmax loss still provides guidance, unlike the weighted sharpness loss).
\end{enumerate}
A weighted sum of these three loss terms constitutes the total loss that is minimized via gradient descent.

\subsubsection{Inference: generating the 6-GP gigamosaics}\label{inference}
Once the CNN that maps z-stacks to height maps is trained, we can apply the CNN to the entire 2.1-TB dataset to generate the 6-GP gigamosaics (Fig. \ref{fig:algorithm}c,d). Specifically, we use a sliding 576$\times$576 window across each 3072$\times$3072 image, looping through the each of the 8$\times$8 lateral scans of each of the 9$\times$6 cameras. We thus read 576$\times$576 z-stacks (with 65 slices each) sequentially from storage, which we feed through the trained CNN to generate height map patches. These height maps are then used to index into the z-stacks to generate the AiF photometric image ($\mathit{AiF}_\mathit{RGB}$). Note that only the green channel is fed into the CNN while all three color channels are used for creating the AiF image patch. Finally, these height map and AiF image patches are homographically dewarped according to the precalibrated camera parameters to enable correct placement of the patches within the gigamosaic. After looping through the entire dataset, we obtain the 6-GP photometric gigamosaic with the coregistered 3D height map. 

As we're accumulating patches, in regions where patches overlap, we have the option to blend, that is, average overlapping pixels (see Supplementary Sec. \ref{blending} for discussion on blending). In Figs. \ref{fig:pin_array}-\ref{fig:painting}, we show the global photometric views using blending and zoom-ins without blending, but we show all 3D height maps with blending.

\begin{figure}
    \centering
    \centerline{\includegraphics[width=\textwidth]{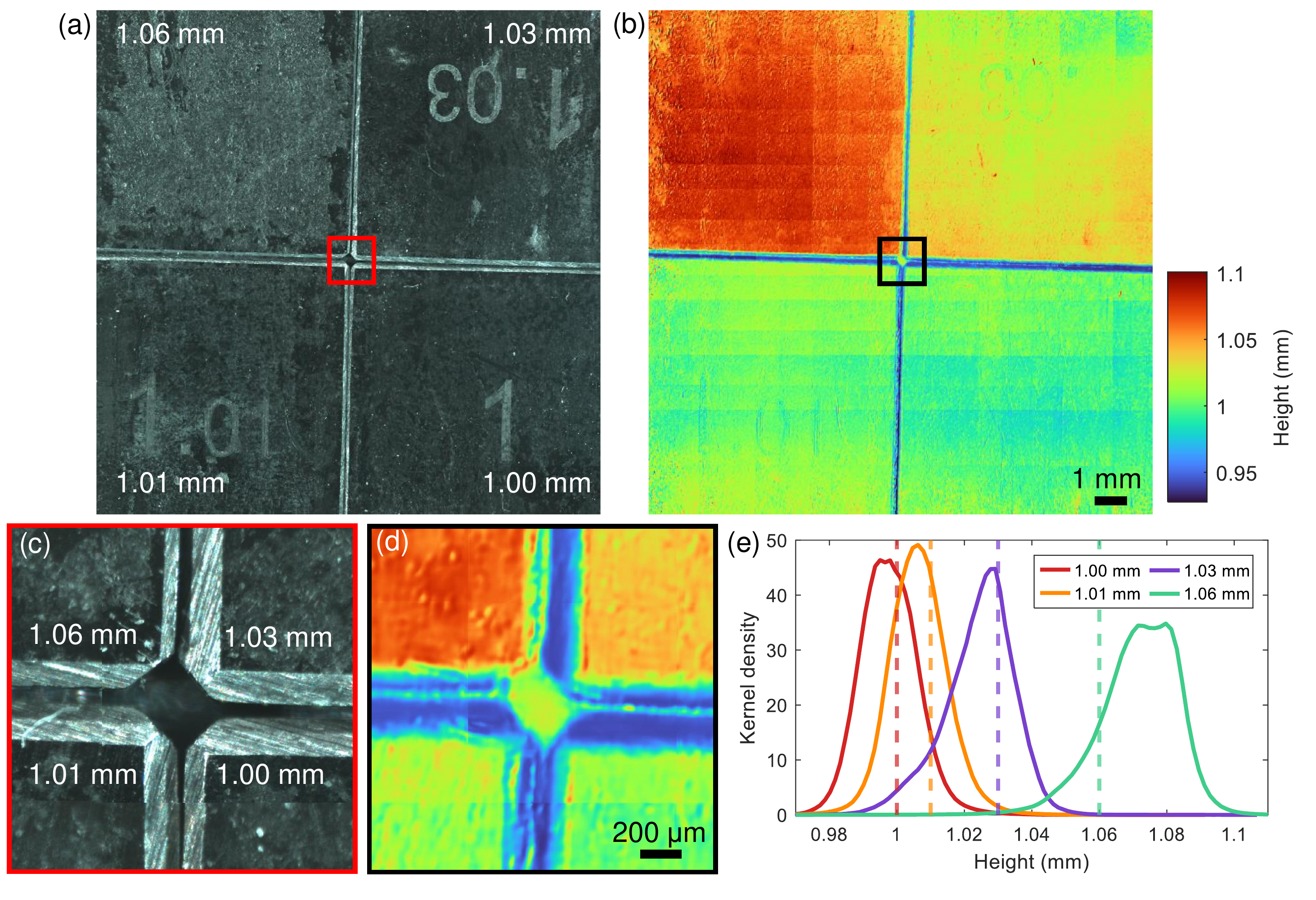}}
    \caption{Axial accuracy and resolution characterization. (a) All-in-focus photometric composite of four gauge blocks with the denoted heights. (b) 3D height reconstruction of the four gauge blocks.(c)-(d) Zoom-ins of (a) and (b). (e) Kernel density estimates of the distribution of the height values of the four gauge blocks. Dotted vertical lines are the ground truth heights.}
    \label{fig:gauge_blocks}
\end{figure}

\section{Results}
\subsection{System characterization}\label{characterization}
We characterized the lateral resolution of STARCAM by imaging a USAF test chart at the center and edge of the FOV of a single camera view (Supplementary Fig. \ref{fig:characterization}(a)). Our system can resolve group 7 element 6, corresponding to a bar width of $\sim$2 \textmu m (or a full-pitch resolution of 4 \textmu m). The DOF of our system is $\sim$0.37 mm, full-width at half maximum (FWHM) (Supplementary Fig. \ref{fig:characterization}(b)). 

To characterize the height accuracy and precision, we applied our method to image four precisely machined (sub-micron accuracy) gauge blocks (Mitutoyo), with heights of 1.000, 1.010, 1.030, and 1.060 mm (Fig. \ref{fig:gauge_blocks}). The axial step size of the z-stack was 2.5 \textmu m. The distributions of the height estimates of the gauge blocks across the imaged area are shown in Fig. \ref{fig:gauge_blocks}(e). The mean height estimates of the gauge blocks were, respectively, 0.998, 1.007, 1.024, and 1.072 mm. Note that since the height map has an arbitrary offset, we set the offset to that which minimizes the mean square error between the means and the ground truths. The accuracy is thus sub-10 \textmu m (mean absolute error = 5.8 \textmu m; root mean square error = 6.8 \textmu m). The precision, quantified by the standard deviation of the height estimates of the gauge block, is approximately 10 \textmu m. Thus, the height estimation accuracy and precision are better than the system DOF.
Finally, note that the gauge blocks have beveled edges, whose lower heights are correctly predicted. Our 3D height map assigns an arbitrary value to the gaps in between the gauge blocks, as nothing comes in focus in the z-stack (Fig. \ref{fig:gauge_blocks}(d)).

\begin{figure}
    \centering
    \centerline{\includegraphics[width=1.4\textwidth]{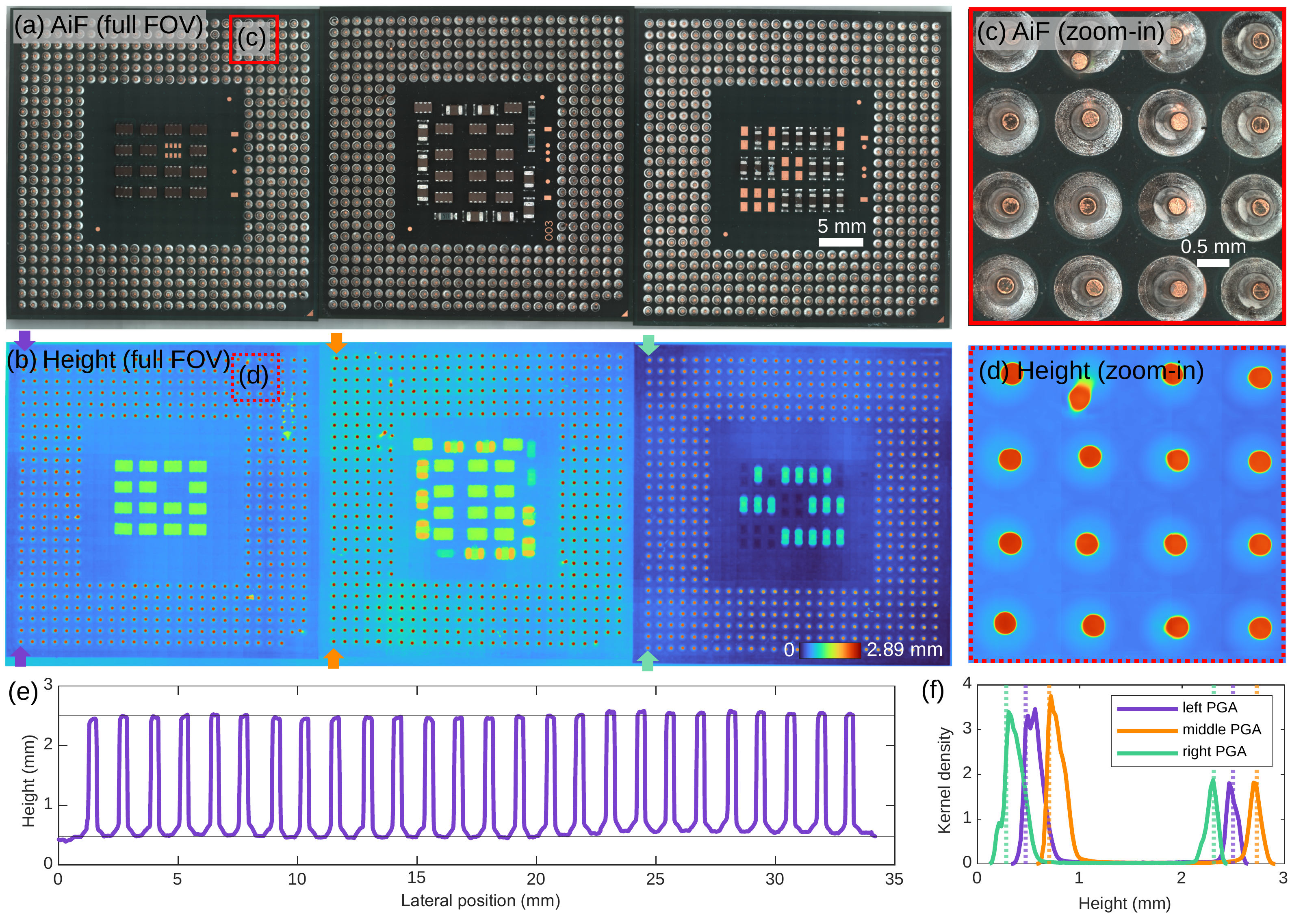}}
    \caption{Three Intel Socket 478 pin grid arrays (PGAs) imaged in parallel. (a) All-in-focus (AiF) photometric gigamosaic. (b) 3D height map. (c) Zoom-in of the region denoted in (a) and the corresponding height map (d). (e) 1D cross section of a row of pins denoted with purple arrows in (b). Horizontal lines are separated by 2.03 mm, the nominal specification. (f) Kernel density estimates of the height distribution of the pins in the rows denoted by the arrows in (b). Vertical dotted lines are separated by 2.03 mm for each PGA.}
    \label{fig:pin_array}
\end{figure}

\begin{figure}
    \centering
    \centerline{\includegraphics[width=1.4\textwidth]{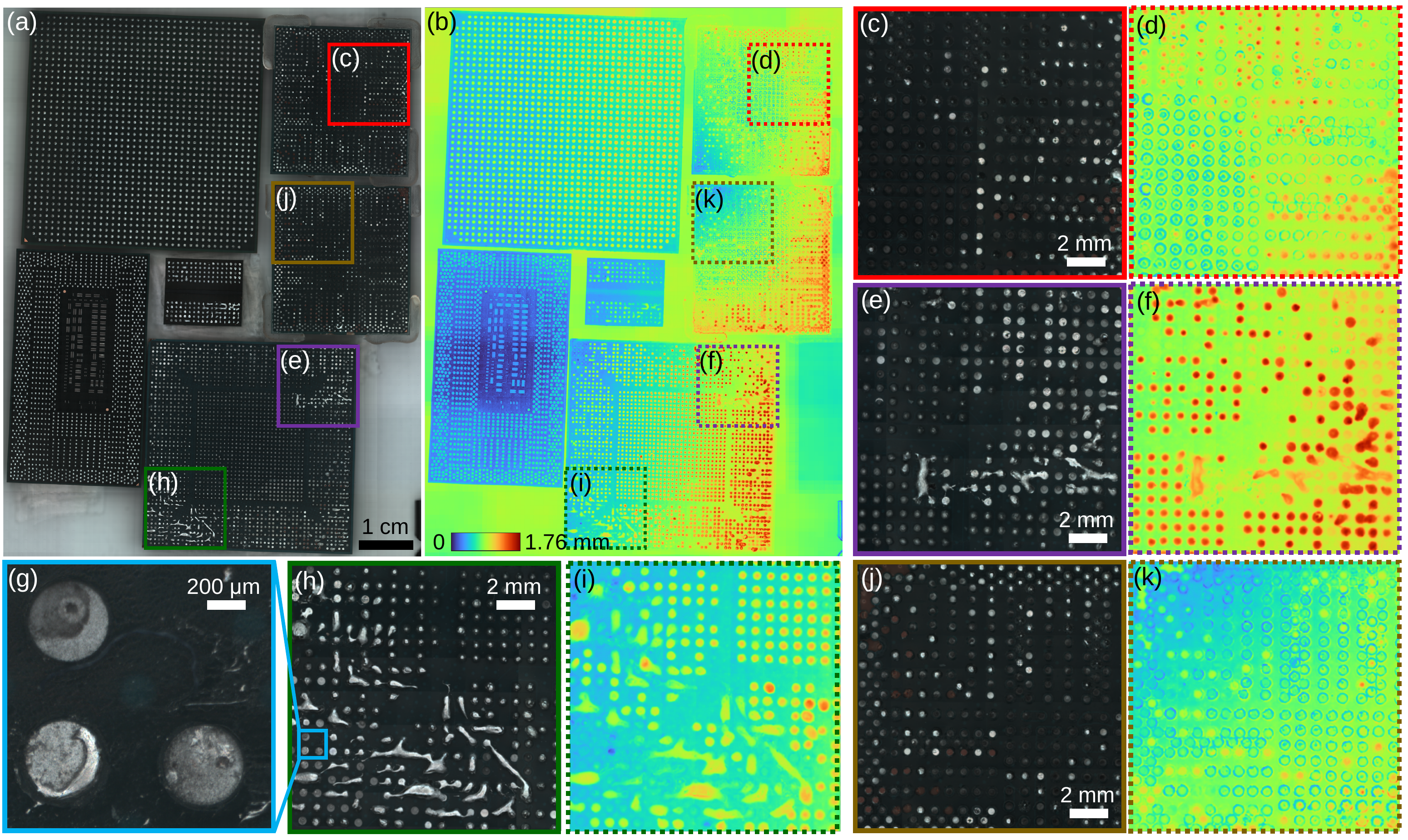}}
    \caption{An assortment of six ball grid arrays (BGAs) imaged in parallel. (a) All-in-focus photometric gigamosaic. (b) 3D height map. (c,e,h,j) Zoom-ins of (a). (d,f,i,k) Zoom-ins of (b). (g) Zoom-in of (h).}
    \label{fig:BGA}
\end{figure}

\begin{figure}
    \centering
    \centerline{\includegraphics[width=1.4\textwidth]{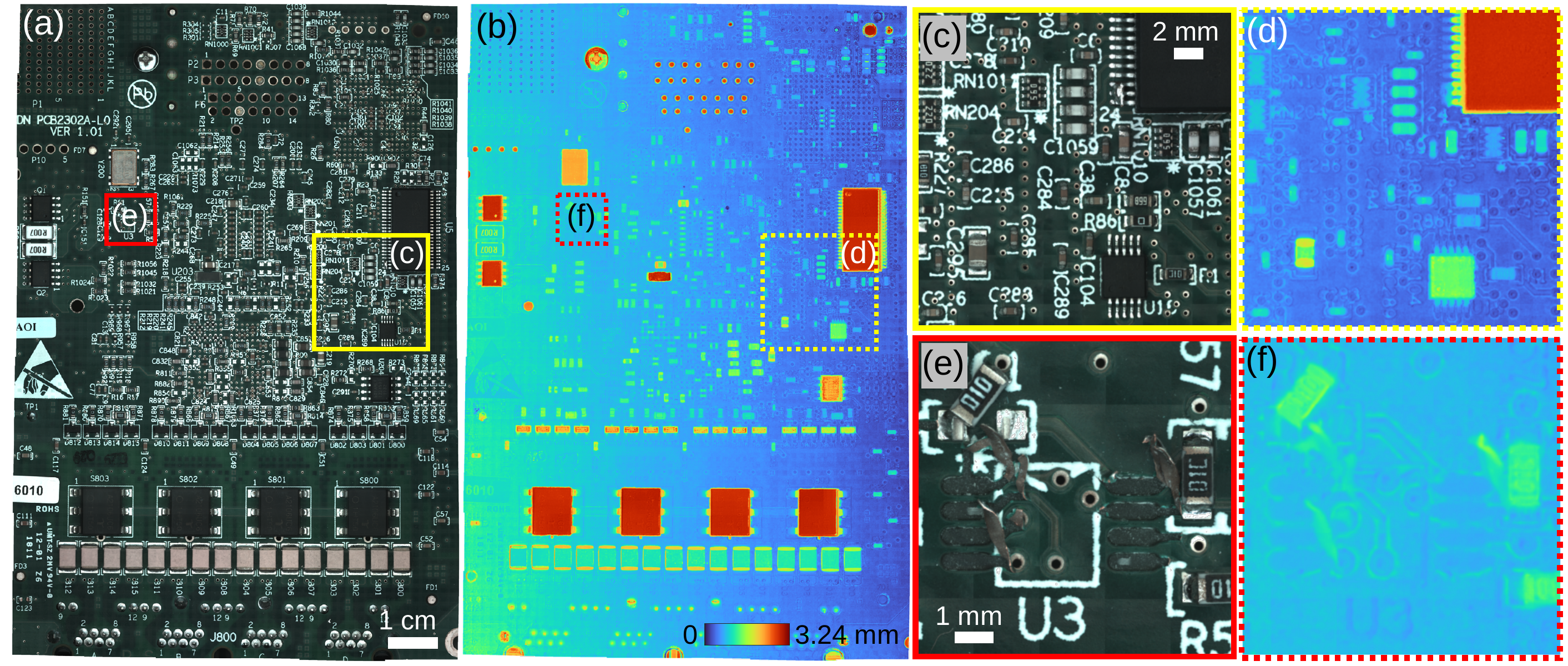}}
    \caption{Printed circuit board (PCB). (a) All-in-focus photometric gigamosaic. (b) 3D height map. (c-f) Zoom-ins of (a,b).}
    \label{fig:PCB}
\end{figure}

\begin{figure}
    \centering
    \centerline{\includegraphics[width=1.2\textwidth]{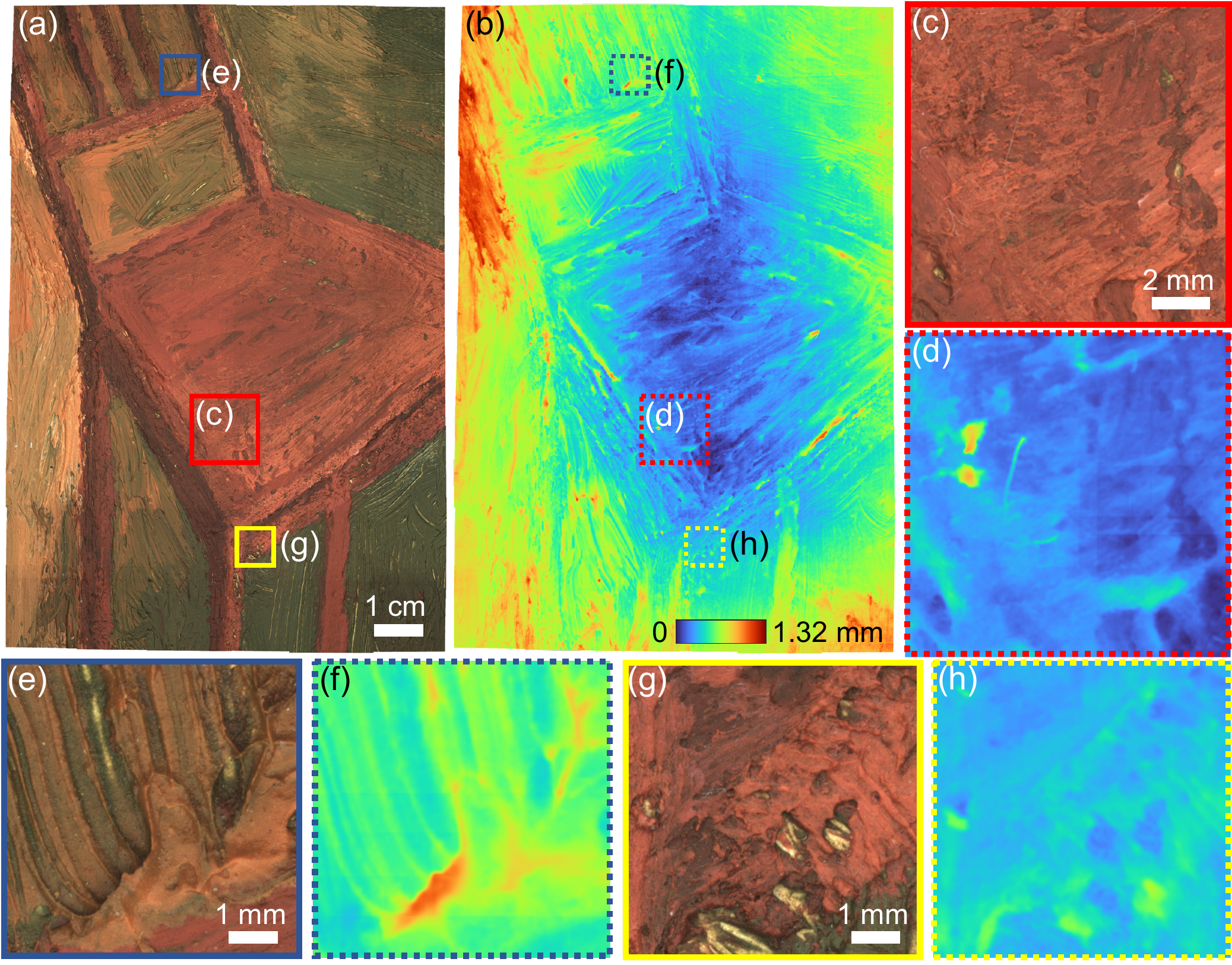}}
    \caption{Oil painting of a chair. (a) All-in-focus photometric gigamosaic. (b) 3D height map.(c-h) Zoom-ins of (a,b) reveal the topographies of different brush strokes.}
    \label{fig:painting}
\end{figure}

\subsection{3D topographic reconstructions of a variety of samples}
We next applied STARCAM to four different extended samples: a pin grid array (PGA) (Fig. \ref{fig:pin_array}), ball grid array (BGA) (Fig. \ref{fig:BGA}), printed circuit board (PCB) (Fig. \ref{fig:PCB}), and oil painting (Fig. \ref{fig:painting}). 

\textbf{Pin grid array.} We imaged three PGAs in parallel, each based on Socket 478 used in Intel's Pentium 4 CPU processors. They each contain an array of 478 pins covering a 35 $\times$ 35 mm\textsuperscript{2} area: 26 $\times$ 26 array of pins with a 14 $\times$ 14 gap in the center, as well as two fewer pins in an outer row (bottom row in Fig. \ref{fig:pin_array}(a)). According to the manufacturing specifications, the pins have a nominal height of 2.03$\pm$0.08 mm. Our $>$110 cm\textsuperscript{2} FOV in principle could have supported at least another three PGAs, with some extra area to spare. Fig. \ref{fig:pin_array}(a-d) show the AiF photometric composite and 3D height map. The height profile for a single row of pins is plotted in Fig. \ref{fig:pin_array}(e), whose heights match the nominal specification of 2.03 mm (horizontal lines). In Fig. \ref{fig:pin_array}(f), kernel density estimates of the height distributions are shown for a row of pins from each PGA (denoted by arrows in Fig. \ref{fig:pin_array}(b)). The two peaks of the bimodal distributions represent the tip of the pins and the base of the chip. Although each PGA has a different height offset, the relative heights of the pins still match the nominal value of 2.03 mm (vertical dotted lines). Note also that some pins are bent away from their normal orientation, which can be seen in the height map.

\textbf{Ball grid array.} We also imaged an assortment of BGA chips (Fig. \ref{fig:BGA}), which consist of arrays of pads with solder balls acting as the interface between integrated circuit chips and PCBs. Some of the chips were forcibly detached, resulting in regions with various heights, depending on whether parts of or the entire solder balls detached. Fig. \ref{fig:BGA} shows the six BGA chips imaged simultaneously, along with several zoom-ins, highlighting the heterogeneity of the damaged solder balls and pads, both in the photometric color images and the 3D height maps. The 3D height map indicates that some of the chips are tilted. The high resolution of our system also allows capture of sub-pad features that would be missed by conventional wide-FOV inspection devices (Fig. \ref{fig:BGA}(g)).

\textbf{Printed circuit board.} We imaged a PCB covering the entire $>$110 cm\textsuperscript{2} FOV of our computational microscope (Fig. \ref{fig:PCB}). The AiF photometric composite reveals fine features at micron-scale resolution of various capacitors, resistors, microchips, and other electronic components across the whole FOV, while the 3D reconstruction reveals their heights. For example, Fig. \ref{fig:PCB}(e,f) shows an example of a detached resistor as well as evidence of a removed microchip (component ``U3''). We note that of all the objects to which we applied our method, this PCB sample posed the most challenges due to the surface texture properties. In Supplementary Sec. \ref{confidence_maps}, we discuss the use of a weighted sharpness (discussed in Sec. \ref{self_supervised}) and confidence maps to address these challenges.

\textbf{Oil painting.} Finally, we applied our method to image a large oil painting, covering the entire $>$110 cm\textsuperscript{2} FOV (Fig. \ref{fig:painting}). The AiF photometric and 3D height map reconstructions reveal not only the textures of different brushstrokes (Fig. \ref{fig:painting}(c-h)), but also dust particles and fibers that have accumulated above the dried paint (Fig. \ref{fig:painting}(c-d)).

\section{Discussion}
We have presented a new large-scale computational 3D topographic microscope, STARCAM, that enables high-SBP imaging (6 GP) of 3D samples at micron-scale resolution over a $>$110 cm\textsuperscript{2} synthetic FOV. We achieve this using a highly parallelized hardware design \cite{harfouche2023imaging}, consisting of 54 cameras along with 3-axis sample scanning to generate a multi-dimensional, multi-TB dataset for each sample, which is then distilled into a 6 GP photometric gigamosaic and a coregistered 3D height map using a CNN-based, self-supervised joint 3D reconstruction and stitching algorithm. The CNN offers a compressed differentiable representation that allows for memory-efficient computational reconstruction across the entire 2.1-TB-per-sample datasets. The self-supervision comes from the stereo overlap redundancy from lateral sample scanning and from image sharpness cues from axial sample scanning. Note that although our method uses sample scanning, we only need to scan laterally across the inter-camera spacing (1.35 cm), and can in principle operate 54$\times$ faster than a conventional single-camera system. We applied our method to a variety of samples, including a PCB, integrated circuit components, and a painting, exemplifying the broad applicability of our method.

While we have demonstrated the potential of STARCAM, there are several avenues for future direction. Currently, besides the diffraction limit and aberrations of the imaging optics of our MCAM, the resolution is further limited by the registration accuracy and the axial step size of the z-stacks. The registration accuracy in turn is currently bottlenecked not by the registration algorithm but by the sample scanning repeatability and accuracy (see Supplementary Sec. \ref{blending}), which can be improved with a more accurate stage and secure sample mounting (or scanning the MCAM instead of the sample). Alternatively or additionally, we could optimize the camera poses jointly with the 3D height maps, as we did previously when the camera poses could not be precalibrated \cite{zhou2021mesoscopic}. The axial step size also affects not only the axial resolution and accuracy, but also the lateral resolution. Currently, we are taking 65 equal steps axially, regardless of the axial scan range, meaning our step sizes in practice were coarser than the empirical results described in Sec. \ref{characterization} and Fig. \ref{fig:gauge_blocks}. Furthermore, if the axial sampling is too coarse, it can result in reduced lateral resolution due to the sample missing the DOF of the objectives. Finally, while our registration model includes the 6 degree-of-freedom camera poses and quadratic radial distortion parameters, we could improve our model by modeling field curvature.

Another possible direction would be to incorporate active patterned illumination \cite{scharstein2003high,yen2006full,geng2011structured,xu2020line,hu2020microscopic} to improve height estimation, especially for low-contrast samples (Supplementary Sec. \ref{confidence_maps}). While it could be a challenge to generate a high-SBP pattern covering the entire $>$110 cm\textsuperscript{2} FOV, a low-resolution structured pattern covering the whole FOV could still be beneficial in combination with the intrinsic sample features. Alternatively, partially coherent off-axis illumination could be used to emphasize edges and small feature for stereo-based height estimation.

In sum, we have developed a high-SBP computational 3D topographic microscope based on a parallelized camera array design and a computational reconstruction and stitching algorithm. Our method can be employed to characterize the 3D topographies of a broad range of extended samples at high resolution, or multiple smaller samples in parallel, with promising applications in accelerating in-line industrial inspection and cultural heritage digitization. Furthermore, our work demonstrates the feasibility of computational imaging with joint, end-to-end optimization across large, multi-TB-per-sample datasets, opening the door to upscaling of other types of computational imaging problems beyond high-SBP topographic microscopy.

\begin{backmatter}
\bmsection{Funding}
Research reported in this publication was supported by the Office of Research Infrastructure Programs (ORIP), Office Of The Director, National Institutes Of Health of the National Institutes Of Health and the National Institute Of Environmental Health Sciences (NIEHS) of the National Institutes of Health under Award Number R44OD024879, the National Cancer Institute (NCI) of the National Institutes of Health under Award Number R44CA250877, the National Institute of Biomedical Imaging and Bioengineering (NIBIB) of the National Institutes of Health under
Award Number R43EB030979, the National Science Foundation under Award
Number 2036439.

\bmsection{Acknowledgments}
We would like to thank Dr. Aurélien Bègue for providing helpful feedback, Prof. Navid Asadi for providing the PCB, and Margaret Aery for painting and providing the oil painting of a chair.

\bmsection{Disclosures}
MH and RH are cofounders of Ramona Optics Inc., which is commercializing the MCAM. MH, MZ, RA, PR, TD, VS, and GH are or were employed by Ramona Optics Inc. when conducting this research. KCZ is a consultant for Ramona Optics Inc. RH is a founder of MIRA Inc., which is applying MCAM technology to art. RA is employed by MIRA Inc. The remaining authors declare no competing interests.

\bmsection{Data Availability Statement}
Data underlying the results presented in this paper are not publicly available at this time, due to the exceedingly large size ($\sim$10 TB), but may be obtained from the authors upon reasonable request.

\bmsection{Code Availability Statement}
The Python code used to generate the gigamosaics and 3D height maps will be provided on Github.

\bmsection{Supplemental document}
See Supplement 1 for supporting content. 

\end{backmatter}

\bibliography{sample}

\newpage

\renewcommand{\thesection}{S\arabic{section}}
\renewcommand{\thetable}{S\arabic{table}}
\renewcommand{\thefigure}{S\arabic{figure}}
\renewcommand\theequation{S\arabic{equation}}
\setcounter{figure}{0}
\setcounter{table}{0}
\setcounter{section}{0}
\setcounter{equation}{0}

\section*{Supplementary Information}
\section{Lateral resolution and depth of field}
\begin{figure}[b]
    \centering
    \includegraphics[width=.7\textwidth]{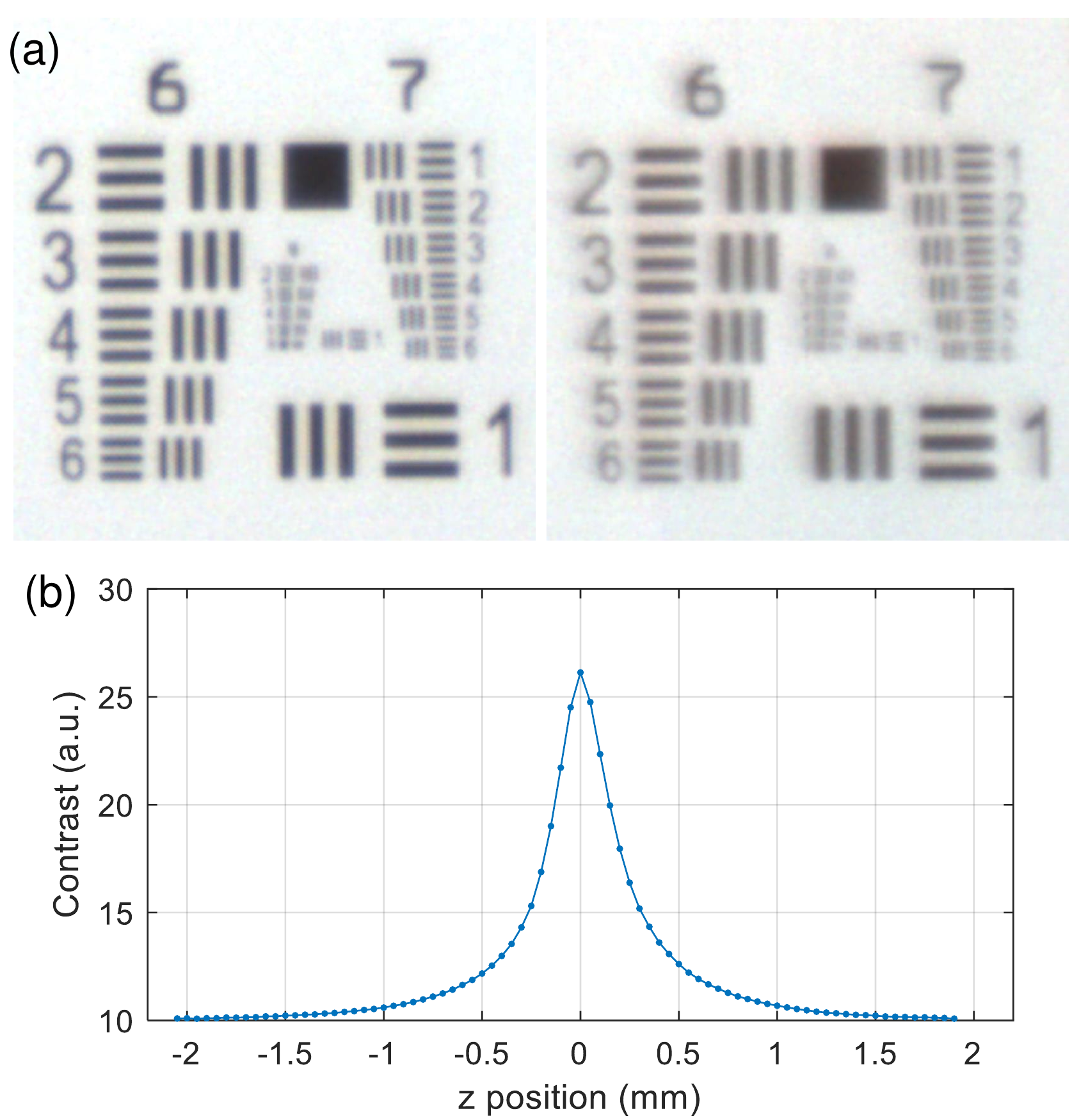}
    \caption{Lateral resolution (a) and DOF (b) characterization of our imaging system. In (a), the left USAF target was placed at the center of the FOV, while the right USAF target was placed at the edge of the FOV.}
    \label{fig:characterization}
\end{figure}
Supplementary Fig. \ref{fig:characterization}(a) shows in-focus images of a USAF target captured by one of the 54 cameras, which are nominally identical. Supplementary Fig. \ref{fig:characterization}(b) characterizes the DOF of our system, based on acquiring a z-stack of a calibration target consisting of white noise printed on paper adhered to a glass substrate. The plot shows the mean image gradient as a function of axial position. This figure is analyzed in Sec. \ref{characterization}.

\section{Sample scan repeatability and calibration errors}\label{blending}
Since we are stitching multiple images into a single composite, seams, or stitching boundaries, may appear and may be geometric (spatial misalignment) or photometric (patch-to-patch intensity or color ratio variation). The former in principle can be pre-calibrated by accurately estimating the camera poses of all 3456 views (54 cameras $\times$ 64 sample scan positions) based on a flat calibration target. The latter, however, in general cannot be fully accounted for due to differing scattering properties of and shadows cast by the sample, especially considering that the sample is scanned relative to the light sources. 

To reduce photometric seams, when we accumulate patches during gigamosaic generation (Sec. \ref{inference}), in regions where patches overlap, we have the option to blend (average overlapped pixels). However, when there are calibration errors leading to imperfect registration and geometric seams, blending can cause blurring and therefore reduce spatial resolution. The reason for imperfect registration is that we fix the calibrated camera poses and distortions during 3D height estimation. Thus, any calibration errors remain uncorrected. These calibration errors do not result from our registration algorithm described in Sec. \ref{camera_calibration}, which contains no visible seams or blurring across the full FOV. Rather, the calibration errors come from the repeatability of the sample scanning, which may be affected by the accuracy, precision, or repeatability 3D sample stage, how rigid the sample during data acquisition, and how stationary the object remains relative to the stage (e.g., if it whole object slips or subcomponents of the object dislodge). This calibration issue can be addressed through a more rigid sample mount, better stage (with specifications better than the expected resolution), or scanning the MCAM instead of the object to reduce the impact of sample idiosyncrasies. 

\section{Accounting for low-contrast objects with confidence maps and weighted sharpness}\label{confidence_maps}
\begin{figure}
    \centering
    \centerline{\includegraphics[width=1.3\textwidth]{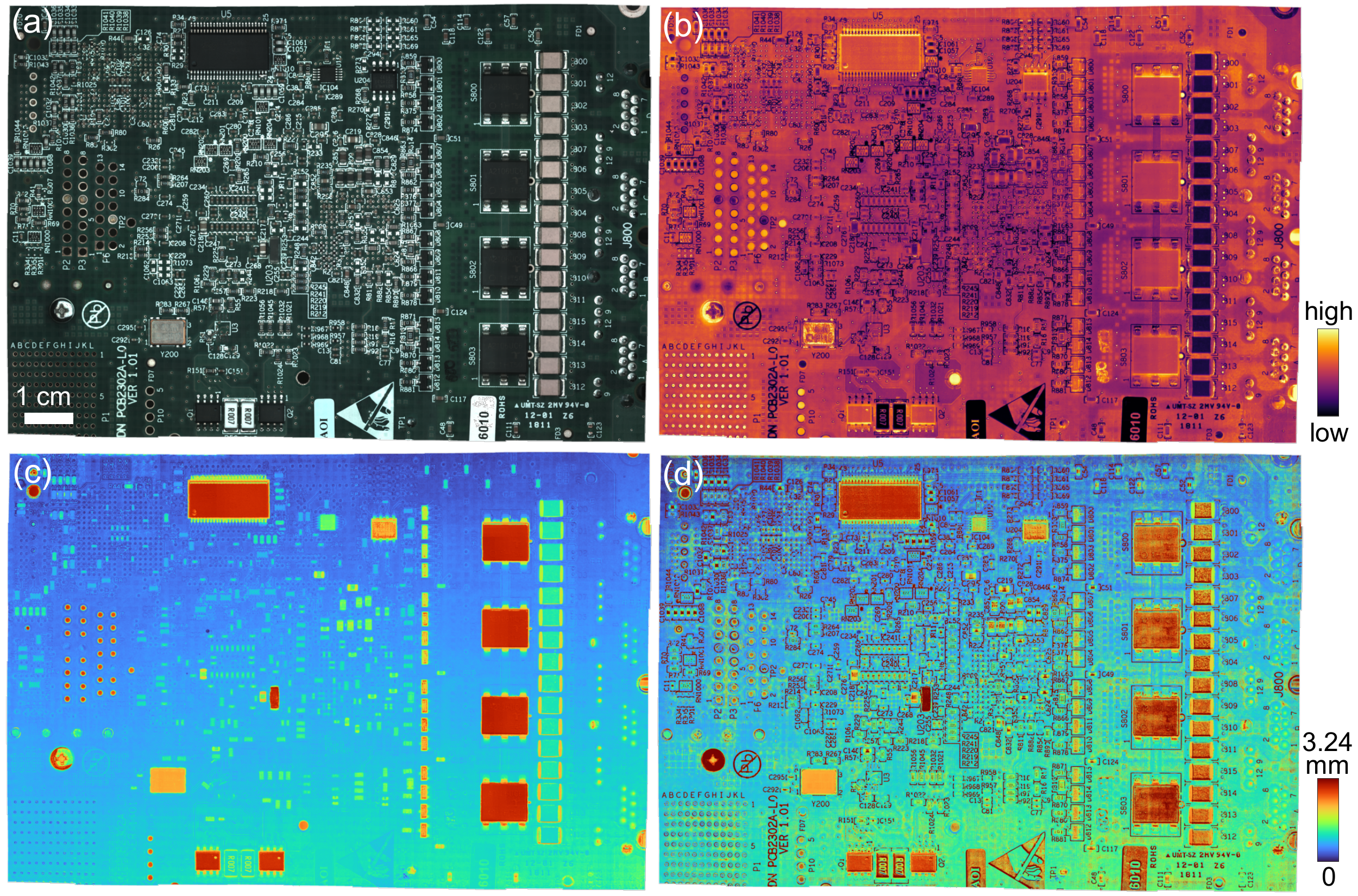}}
    \caption{PCB analysis. (a) All-in-focus photometric gigamosaic. (b) Confidence map, based on max sharpness across the z-stack dimension. (c) 3D height map reconstruction. (d) 3D height map reconstructed using only argmax across the z-stack dimension.}
    \label{fig:PCB_supplement}
\end{figure}
Both stereo- and sharpness-based methods require the sample to have fine texture or features. For stereo methods, these features are registered from different views, while for sharpness methods, these features must have different appearances when translated to different axial positions. Thus, in regions of the sample that are low contrast, the reconstruction may result in larger errors. To account for this, we introduce two strategies. The first is the use of a weighted sharpness loss (Eq. \ref{weighted}), described in Sec. \ref{self_supervised}, with the idea that regions of low sharpness (or low contrast) should contribute less or not at all to the the total loss. The other strategy is to compute a coregistered confidence map, based on the max sharpness values across the z-stack dimension. Fig. \ref{fig:PCB_supplement}(b) shows that specific PCB components (e.g., the row of 16 components on the righthand side) as well as white text, lines, and stickers are particularly low-contrast, thus cautioning the user that the height values may be inaccurate. We can see evidence of this in the argmax-only height map in  Fig. \ref{fig:PCB_supplement}(d), which erroneously assigns inflated height values to the white text and lines, unlike the height map generated by the full version of our method that also incorporates the weighted sharpness loss and stereo (Fig. \ref{fig:PCB_supplement}(c)). Thus, the confidence map can be reported alongside our reconstructions to safeguard or warn the user when parts of the sample of interest are low contrast, and therefore whose height estimates may be inaccurate.

\begin{figure}
    \centering
    \centerline{\includegraphics[width=1.2\textwidth]{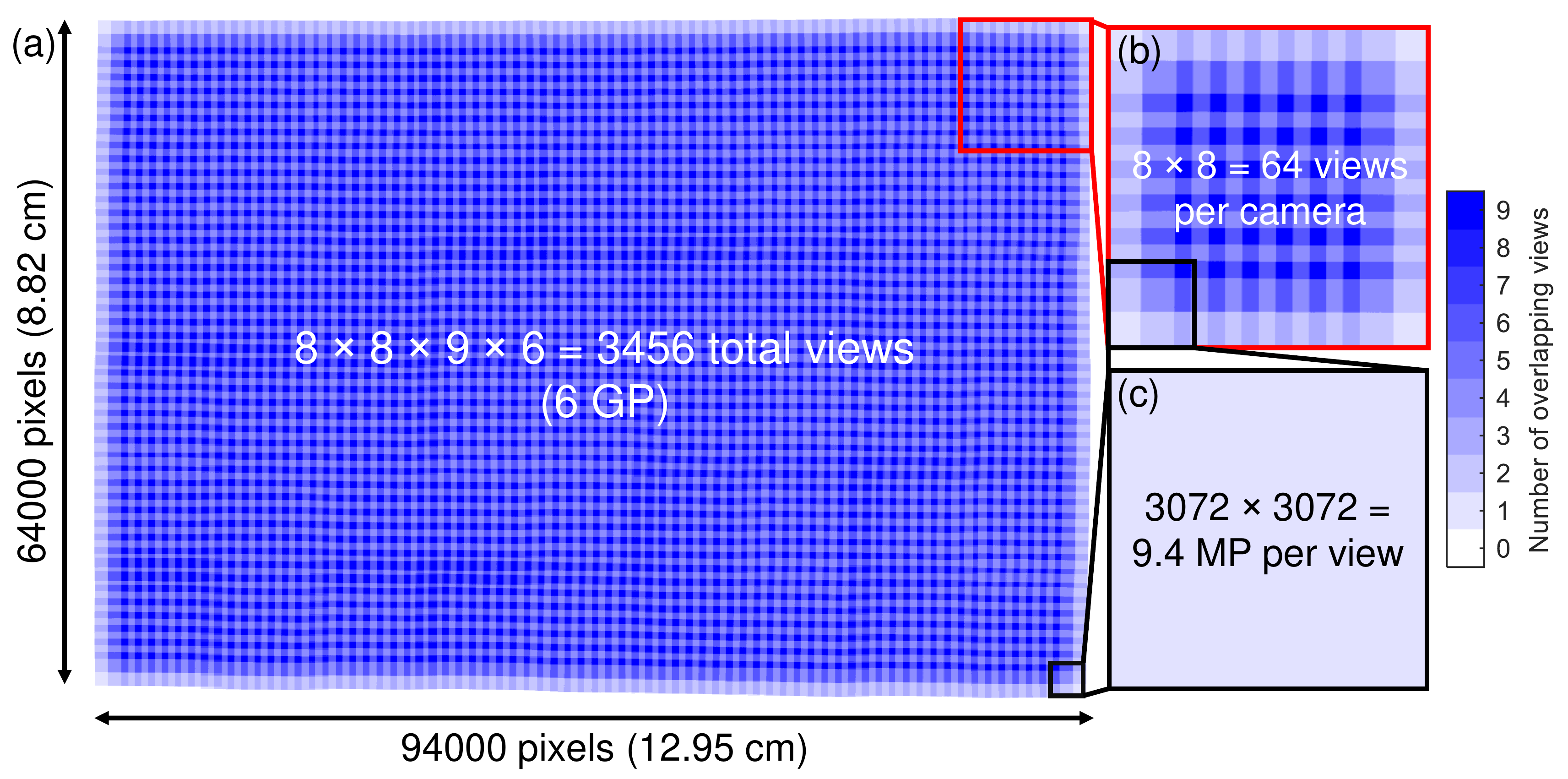}}
    \caption{Overlap maps, showing how many times (0-9) each point in the FOV. (a) The overlap map for all 3456 camera views, based on the joint camera calibration from imaging a flat reference target. (b) The overlap map for one 8$\times$8 scan from a single camera. (c) Each view contains over 9 MP.}
    \label{fig:calibration_overlap}
\end{figure}

\end{document}